\RequirePackage{amsmath}
\documentclass[prb,aps,twocolumn,natbib,showkeys,showpacs] {revtex4}

\usepackage[dvips]{graphicx}
\usepackage{epsfig}
\usepackage{amssymb}

\DeclareMathOperator{\sech}{sech}
\DeclareMathOperator{\sgn}{sgn}
\DeclareMathOperator{\erfc}{erfc}

\begin{document}

\title{Soliton scattering as a measurement tool for weak signals}

\author{I. I. Soloviev$^{1,2,3}$}
\email[]{igor.soloviev@gmail.com}
\author{N. V. Klenov$^{4,1,2,3}$}
\author{A. L. Pankratov$^{5,6,7}$}
\email[]{alp@ipmras.ru}
\author{L. S. Revin$^{5,6,7}$}
\author{E. Il'ichev$^{8}$}
\author{L. S. Kuzmin$^{7,9,1}$}
\affiliation{$^{1}$Lomonosov Moscow State University Skobeltsyn
Institute of Nuclear Physics, 119991, Moscow, Russia}
\affiliation{$^{2}$Lukin Scientific Research Institute of Physical
Problems, Zelenograd, 124460, Moscow, Russia}
\affiliation{$^{3}$Moscow Institute of Physics and Technology, State
University, Dolgoprudniy, Moscow region, Russia}
\affiliation{$^{4}$Physics Department, Moscow State University,
119991, Moscow, Russia} \affiliation{$^{5}$Institute for Physics of
Microstructures of RAS, Nizhny Novgorod, 603950, Russia}
\affiliation{$^{6}$Lobachevsky State University of Nizhni Novgorod,
Nizhny Novgorod, 603950, Russia} \affiliation{$^{7}$Nizhny Novgorod
State Technical University n.a. R.E. Alekseev, Nizhny Novgorod,
603950, Russia} \affiliation{$^{8}$Leibniz Institute of Photonic
Technology, D-07702 Jena, Germany}\affiliation{$^{9}$Chalmers
University of Technology, SE-41296 Goteborg, Sweden}

\date{\today}



\begin{abstract}
We have considered relativistic soliton dynamics governed by the
sine-Gordon equation and affected by short spatial
inhomogeneities of the driving force and thermal noise.
Developed analytical and numerical methods for calculation of
soliton scattering at the inhomogeneities allowed us to examine the
scattering as a measurement tool for sensitive detection of polarity
of the inhomogeneities. We have considered the superconducting fluxonic
ballistic detector as an example of the device in which the soliton
scattering is utilized for quantum measurements of superconducting
flux qubit. We optimized the soliton dynamics for the measurement
process varying the starting and the stationary soliton velocity as
well as configuration of the inhomogeneities. For experimentally
relevant parameters we obtained the signal-to-noise ratio above 100
reflecting good practical usability of the measurement concept.
\end{abstract}

\maketitle

\section{Introduction}

Solitary waves (named solitons) preserving their shape due to a
strong nonlinear interaction with the medium in which they propagate
are well known from macroscopic to microscopic scales \cite{MK}. One
of the equations having soliton solution is the sine-Gordon (SG)
one. This equation describes a variety of nonlinear systems
\cite{MK, WLLU, snig, Ust, NFP, KS, pskm, dna, dna2, dna3, dna4,
Hut,VGS} among which are superconducting devices devoted for
information receiving and processing \cite{SDA,Miki}, including
quantum schemes \cite{qubit,WKU,FSSU,BBPH}. For read out the last
ones, the well known high sensitivity of superconducting detectors
\cite{CB} can be conjugated with evanescent back-action on the
measured object using a special readout concept, e.g. the ballistic
readout \cite{A}. Operation principle of the ballistic readout is
based on ability of a measured object to affect transport of
particles by inducing scattering potential for them, which is
similar to the idea of the Rutherford experiments. Due to their
inherent particle-like stability joint with a wave nature, the
solitons (which are the fluxons in superconducting circuits) are the
natural candidates for the role of particles in the scheme. Such
fluxonic detector was proposed \cite{ARS}, studied \cite{HFSIS,
FSSK, F, SKBPK} and tested experimentally \cite{FSSU,FSWBU}. It has
been argued that all types of measurements known in quantum
mechanics can be realized using this approach. Namely, the
measurements can be done in single-shot \cite{SKBPK}, weak
continuous (in some literature called ``non-projective'')
\cite{MQ,BVT} and nearly non-demolition \cite{ARS} regimes that have
grown an interest in the research motivated by possibility of
exploration of such fundamental scientific concepts as ``wave
function collapse'' and decoherence.

\begin{figure}[b]
\resizebox{1\columnwidth}{!}{
\includegraphics{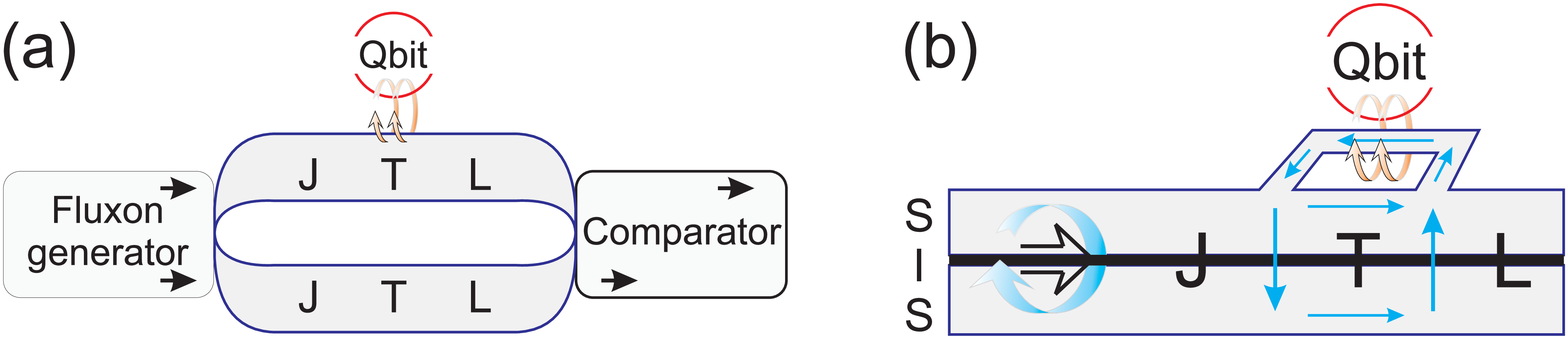}}
\caption{(a) Fluxonic ballistic detector scheme. (b) Transformation
of the qubit magnetic field into the current dipole in the JTL (blue
arrows show currents induced by the field and the fluxon currents).
Black arrows show fluxons. S denotes superconductor and I -
isolator.} \label{Fig1}
\end{figure}

One option of the considered measurement scheme for detection of
weak magnetic field (which can be a flux qubit field) is shown in
Fig.~\ref{Fig1}a. In this interferometric scheme a couple of fluxons
simultaneously propagate through a couple of identical Josephson
transmission lines (JTLs). The measured object, being coupled with
one of the JTLs, introduces its weak magnetic field into this JTL,
where it is transformed to the current dipole (the dipole of the
driving force affecting the soliton motion) as shown in
Fig.~\ref{Fig1}b. Fluxon scattering at this current dipole leads to
deviation of its propagation time from the ones of the fluxon
propagating through the reference (uncoupled) JTL. The time
difference can be detected at the output comparison circuit if its
magnitude is well above the noise level. The measurements, based on
soliton scattering, can be realized also in frequency domain as it
was done in experimental works \cite{FSSU,F,FSWBU}.

Experimental results revealed an importance of accounting for
relativistic aspects of fluxon dynamics in estimation of the
detector response, while theoretical works devoted to the detector
mainly considered non-relativistic regime with stationary fluxon
velocity \cite{HFSIS,FSSK} because of mathematical difficulties. The
non-relativistic approach was also traditionally used for estimation
of noise effect on fluxon dynamics in digital superconducting
circuits \cite{RL, Ter, Ter2}. However, the recent works
\cite{SKPIK,PGK} have shown that such relativistic effects as Lorenz
contraction of the fluxon shape and change of its effective mass
drastically affect the noise properties of the system.

In this paper we develop analytical and numerical methods for
modeling of the soliton scattering at inhomogeneities of the driving
force and accounting for the thermal fluctuations, comprising
consideration of relativistic regime. The developed approaches allow
us to calculate the signal-to-noise ratio (SNR) of the measurement
procedure based on soliton scattering. For particular example of the
original fluxonic ballistic detector scheme \cite{ARS} we find
dependences of the SNR as function of the driving force as well as
the location of its inhomogeneities induced by a measured object.
Tuning of the detector parameters allows us to obtain the SNR values
above 100 that proves practical applicability of the considered
measurement concept.

\section{Calculation of soliton scattering dynamics}

Let us consider the SG equation describing a JTL. For
superconducting phase difference $\phi$ it can be written in the
following form:
\begin{equation}
{\phi}_{tt}-{\phi}_{xx}+\sin(\phi)=-\alpha{\phi}_{t} + i + i_f(x,t)
+ i_s(x). \label{SG}
\end{equation}
Here the first two terms ($-\alpha\phi_t$ and $i$) in the right-hand
side of the equation (\ref{SG}) represent the energy dissipation due
to tunneling of normal electrons across the barrier and the overlap
bias current density providing the energy input. The next two
terms ($i_f(x,t)$ and $i_s(x)$) account for the thermal fluctuations
and scattering inhomogeneity of the bias current.

The components of the current densities $i$, $i_f$ and $i_s$ are
normalized to the critical current density $J_c$. The space
coordinate $x$ and the time $t$ are normalized to the Josephson
penetration length $\lambda _{J}$ and to the inverse plasma
frequency $\omega_{p}^{-1}$, respectively;
$\alpha={\omega_{p}}/{\omega_{c}}$ is the damping coefficient,
$\omega_p=\sqrt{2eI_c/\hbar C}$, $\omega_{c}=2eI_cR_{N}/\hbar$,
$I_c$ is the critical current, $C$ is the JTL capacitance, $R_N$ is
the normal state resistance.  The noise correlation function is:
$\left<i_f(x,t)i_f(x',t')\right>=2\alpha\gamma \delta
(x-x^{\prime})\delta (t-t^{\prime})$, where $\gamma = I_{T} /
J_{c}\lambda_J$ is the dimensionless noise intensity
\cite{Fed,Fed2}, $I_{T}=2ekT/\hbar$ is the thermal current, $e$ is
the electron charge, $\hbar$ is the Planck constant, $k$ is the
Boltzmann constant and $T$ is the temperature. If the scattering
inhomogeneity has the width much less than the fluxon characteristic
size $\lambda _{J}$, the corresponding term can be expressed as
$i_s(x) = \mu\delta(x-x_c)$, where $\mu$ is the amplitude and $x_c$
is the central coordinate of the inhomogeneity.

\subsection{Analytical approach.}

Analytical description of the soliton scattering dynamics can be
developed if all the perturbation terms in the equation (\ref{SG})
are small: $\alpha, i, i_f, i_s \ll 1$. In this case one can use the
collective coordinate perturbation theory developed by McLaughlin
and Scott \cite{McS} to obtain the system of nonlinear differential
equations for the soliton velocity $u$ and its central coordinate
$X$ (the details of the calculations are summarized in the
Appendix):
\begin{subequations} \label{duX}
\begin{multline}
\frac{du}{dt} = -\alpha u(1-u^2) - \frac{1}{4}[\pi i +
\xi(t)](1-u^2)^{3/2}\\
-\frac{1}{4}(1-u^2)\mu\sech(\theta),
\end{multline}
\begin{equation}
\frac{dX}{dt} = u - \frac{1}{4} u \sqrt{1-u^2}\mu\theta
\sech(\theta),
\end{equation}
\end{subequations}
where $\theta = (X - x_c)/\sqrt{1-u^2}$, the velocity $u$ is
normalized to the Swihart velocity $c = w_p \lambda_J$, $X$ is
normalized to $\lambda_J$, and the noise intensity \cite{JKMMSS}
$\left<\xi(t)\xi(t')\right> = \alpha\gamma(1-u^2)^{-1/4}\delta(t -
t')$.

This system is too complex to be solved directly. However, one can
find the desired dependences successively considering the scattering
and the noise effect as perturbations to the solution governed by
constant energy gain and loss ($i$ and $\alpha$). This solution (for
$\mu, \xi = 0$ in the system (\ref{duX})) describing the soliton
velocity relaxation process conditioned by $i/\alpha$ ratio is as
follows:
\begin{subequations} \label{uX0}
\begin{equation}
u^{rel}(t) = \sgn(p)\left(1 + p^{-2}\right)^{-1/2},
\end{equation}
\begin{multline}
X^{rel}(t) = u_{st}(t-t_0)+\frac{u_{st}\ln(A_1) - \ln(A_2)}{\alpha}
+C,
\end{multline}
\end{subequations}
where
\begin{equation} \label{p}
p = \left(\beta +
u^{rel}(t_0)/\sqrt{1-u^{rel}(t_0)^2}\right)e^{-\alpha
(t-t_0)}-\beta,
\end{equation}
is the soliton momentum,
\begin{multline}
A_1 = p\left(\sqrt{\beta^2+1}\sqrt{1+p^{-2}}\sgn(p)-\beta\right) +
1,\nonumber \\
A_2 = p\left(\sqrt{1+p^{-2}}\sgn(p)+1\right), \nonumber
\end{multline}
$C$ is constant
\begin{multline}
C=X^{rel}(t_0) - \frac{1}{\alpha}\biggl(u_{st}\ln{\textstyle\left[
\frac{\sqrt{\beta^2+1}-\beta
u^{rel}(t_0)}{\sqrt{1-u^{rel}(t_0)^2}} +1\right]}\\
-\ln{\textstyle\left[\frac{1+|u^{rel}(t_0)|}{\sqrt{1-u^{rel}(t_0)^2}}\right]}\sgn(u^{rel}(t_0))\biggr),\nonumber
\end{multline}
$u_{st}$ is the soliton stationary velocity
\begin{equation} \label{ust}
u_{st} = -\sgn(\beta)(1+\beta^{-2})^{-1/2},
\end{equation}
parameters $t_0$, $X^{rel}(t_0)$, $u^{rel}(t_0)$ are the starting
conditions, $\beta = \pi i/4 \alpha$. From equation (\ref{p}) it is
seen that the soliton velocity relaxation rate is determined by the
damping.

Next, we account for the perturbation provided by the scattering
assuming the ballistic regime: $i, \alpha, \xi = 0$. Approximate
solution of the system (\ref{duX}) in this case has the form:
\begin{subequations} \label{uX1}
\begin{multline}
u^{sc}(\theta) = \sgn(u_0) \\ \times\sqrt{1 - \frac{1-u_0^2}{\left[1
- \frac{\mu}{2}\sqrt{1 -
u_0^2}\left(\arctan\left[\tanh\left(\frac{\theta}{2}\right)\right] +
c_{\pm\infty}\right)\right]^{2}}},
\end{multline}
\begin{multline}
X^{sc}(\theta) = \\
\frac{\theta \sqrt{1-u_0^2}}{1 - \frac{\mu}{2}\sqrt{1 -
u_0^2}\left(\arctan\left[\tanh\left(\frac{\theta}{2}\right)\right] +
c_{\pm\infty}\right)} + x_c,
\end{multline}
\end{subequations}
where $c_{\pm\infty}$ is constant corresponding to solutions for
incident ($c_{-\infty}=\pi/4$) and scattered ($c_{+\infty}=-\pi/4$)
soliton for $u_0 > 0$, and vice versa for the negative velocity $u_0
< 0$.

For an incident soliton, which ballistically propagates with
velocity $u_0$, the scattering provides a step of the velocity
\begin{equation} \label{ustep}
u_0 \rightarrow \sgn(u_0)\sqrt{1 -
\frac{1-u^2_0}{\left(1-\sgn(u_0)\frac{\mu\pi}{4}\sqrt{1-u^2_0}\right)^2}}
\end{equation}
centered at $\theta = 0$. According to the equation (\ref{uX1}b)
this step appears as a bend on the coordinate dependence
\begin{equation} \label{Xstep}
X \rightarrow \frac{X -
x_c}{1-\sgn(u_0)\frac{\mu\pi}{4}\sqrt{1-u^2_0}} + x_c.
\end{equation}

Soliton velocity relaxation to its stationary value can be taken
into account by using the solution (\ref{uX0}) in the system
(\ref{uX1}): $u_0 = u^{rel}$ and $\theta = (X^{rel} -
x_c)/\sqrt{1-(u^{rel})^2}$. The moment of the scattering $t^{sc}$
then can be estimated from the equation $X^{rel}(t^{sc}) = x_c$, so
\begin{equation}
t^{sc} = \int_{t_0}^\infty H(u^{rel}(t)) - \sgn(u^{rel}(t))
H(X^{rel}(t) - x_c) dt + t_0,
\end{equation}
where $H(x)$ is the Heaviside step function.

Since the scattering perturbs the relaxation dynamics of an incident
soliton, the solution governed by the initial starting conditions
$t_0$, $X^{rel}(t_0)$, $u^{rel}(t_0)$ can be used in the system
(\ref{uX1}) only up to the time $t^{sc}$ (for $t < t^{sc}$). After
this time ($t \geq t^{sc}$) one should use the solution for a
scattered soliton with appropriate starting conditions: $t^{sc}$,
$X^{rel}(t^{sc}) = x_c$ and $u^{rel}(t^{sc})$ equal to the shifted
velocity defined by the right-hand side of the expression
(\ref{ustep}), where $u_0 = \lim_{t\to t^{sc}}u^{rel}(t)$ is adopted
from the incident soliton solution for crosslinking.

To account for the effect of noise, we can consider the soliton as a
massive Brownian particle but with the time dependent noise
intensity \cite{PGK}. We omit the terms with $\mu$ in the system
(\ref{duX}) and assume that the velocity in the factors that reflect
the relativistic effects ($(1-u^2)$ and $(1-u^2)^{3/2}$) does not
significantly fluctuate in the low noise limit, so $u$ is
substituted for the found $u^{sc}(t)$ there. For further
simplification we consider some fixed relativistic decrease of the
damping, substituting $u^{sc}(t)$ in the factor $(1-(u^{sc})^2)$ in
front of the damping term for the average velocity
$\left<u^{sc}(t)\right>$ (which derivation will be outlined below),
so the effective damping is $\alpha^* =
\alpha(1-\left<u^{sc}\right>^2)$. Using these approximations, for
Gaussian noise $\xi(t)$ we find the variance $D(t)$ and the
corresponding probability $P(t)$ to find the soliton inside the
segment of the length $L$ in the following form:
\begin{multline} \label{D}
D(t) = \frac{\gamma}{4 \alpha^*}\int_{t_0}^t \left(1 - 2
e^{-\alpha^* t'} + e^{-2\alpha^* t'}\right)
\\
\times\left[1 - (u^{sc}(t'))^2\right]^{5/2}dt',
\end{multline}
\begin{equation} \label{P}
P(t) = 1 - \frac{1}{2}\erfc\left[\left(L -
X^{sc}(t)\right)/\sqrt{2D(t)}\right].
\end{equation}
These equations allow obtaining the mean soliton propagation time
$\tau$ through the segment and its standard deviation $\sigma$
(jitter) using the notion of the integral relaxation time
\cite{IntRel}:
\begin{equation} \label{ts}
\setcounter{MaxMatrixCols}{20}
\begin{matrix}
{\tau = \int_{t_0}^\infty P(t)dt,} && {\sigma =
\sqrt{2\int_{t_0}^\infty tP(t)dt - \tau^2}}.
\end{matrix}
\end{equation}
In the limit $D(t)\rightarrow 0$ the equations (\ref{P}), (\ref{ts})
serve for estimation of the propagation time without noise and
corresponding average soliton velocity $\left<u^{sc}\right>$, which
in turn is used for calculation of the effective damping $\alpha^*$.

Finally, the SNR of the measurement process based on soliton
scattering can be calculated. For example, if the measurement
implies comparison between soliton propagation times with
$\tau_\mu$ and without $\tau_0$ scattering, then the SNR is:
\begin{equation} \label{SNR}
SNR = \frac{|\Delta\tau|}{\sigma_\Sigma} =\frac{|\tau_\mu -
\tau_0|}{\sqrt{\sigma_\mu^2 + \sigma_0^2}},
\end{equation}
where $\sigma_{\mu,0}$ correspond to the mean times $\tau_{\mu,0}$.

\subsection{General method.}

The described analytical approach can be generalized for any number
of inhomogeneities of the driving force. For example, the scattering
at a dipole can be considered as two successive scatterings at
inhomogeneities spread over a distance of the dipole width $d =
x_{c2} - x_{c1}$ with amplitudes of the opposite sign $\mu_2 = -
\mu_1$. To find solution in this case one should first obtain
$u^{sc}_1$, $X^{sc}_1$ using $\mu_1$ and $x_{c1}$ in the system
(\ref{uX1}) and then use these equations (\ref{uX1}) again with
$\mu_2$, $x_{c2}$, using previously obtained $u^{sc}_1$, $X^{sc}_1$
instead of $u^{rel}$, $X^{rel}$ for incident soliton solution.

If the scattering can not be considered as a perturbation (because
of high scattering amplitude or since the soliton motion can not be
considered as ballistic), one should proceed with numerical
calculation of the system (\ref{duX}) in which the last terms should
be substituted in general for $-(1-u^2)\sum_n
\mu_n\sech(\theta_n)/4$ and $-u\sqrt{1-u^2}\sum_n
\mu_n\theta_n\sech(\theta_n)/4$ in the equations for $u$ and $X$
($n$ is the inhomogeneity number). To speed up the calculations, the
fluctuational term can be omitted (if $\xi \ll 1$) in the numerical
evaluations of $u(t)$ and $X(t)$. The effect of noise can be
accounted further as it is described above, using equations
(\ref{D})-(\ref{SNR}).

At last, if the terms on the right-hand side of the SG equation
(\ref{SG}) are not small, this equation should be calculated
numerically itself. It is useful then to substitute the delta
function in the scattering term for some smoother one, e.g.
hyperbolic secant \cite{FSSU}: $i_s(x) = \sum_n
\mu_n\delta(x-x_{cn}) \approx \sum_n \mu_n\sech[(x-x_{cn})/a_n]/\pi
a_n$, where $a_n$ characterizes the width of the scattering
inhomogeneity. The mean soliton propagation time and its standard
deviation can be obtained by averaging over ensemble of
realizations.

\section{Soliton scattering as a measurement tool of the fluxonic ballistic detector}

\begin{figure}[t]
\resizebox{1\columnwidth}{!}{
\includegraphics{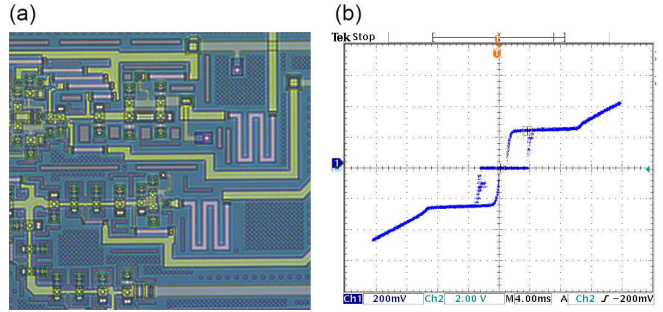}}
\caption{(a) Photo of the experimental sample fragment with digital
superconducting circuits of the fluxon generator block. (b)
Current-voltage characteristic of serial array containing 10 test
Josephson junctions. The current scale (abscissa) is 100~$\mu$A/div,
the voltage scale (ordinates) is 10~mV/div.} \label{Fig2}
\end{figure}

Let us consider the original fluxonic ballistic detector scheme (see
Fig.~\ref{Fig1}) to study the measurements based on soliton
scattering. For verification of the presented theoretical approaches
we compare their results, and furthermore design the superconducting
schemes for measurements of the detector time response and its
jitter. The designs are intended for fabrication by FLUXONICS
foundry \cite{FLUXONICS}. Fragment of one of the fabricated samples
is shown in Fig.~\ref{Fig2}a.

While our measurements are in progress, we have estimated parameters
of Josephson junctions which are necessary for calculations. Typical
current-voltage characteristic of a serial array containing 10 test
junctions is presented in Fig.~\ref{Fig2}b. According to these data,
the junction quality is $R_J/R_N \simeq 20$ (where $R_J$ is the
subgap resistance) and the damping at 4~K temperature is $\alpha
\simeq 0.2$. For experimental temperatures $T \geq 50$~mK we expect
a decrease in the damping value by one order \cite{HFSIS} down to
$\alpha \simeq 0.02$.

Our estimation gives the value of the normalized noise intensity
$\gamma = 10^{-5}$ at 50~mK temperature. However to speed up the
numerical calculations, we mainly used the value $\gamma = 10^{-3}$.
This value is still much smaller than values of the other
coefficients in the SG equation, so the noise effect remains weak.
One should note that the time jitter and the SNR scale accurately as
$\sigma_\Sigma \sim \sqrt{\gamma}$ and $SNR \sim 1/\sqrt{\gamma}$,
respectively, that has been proven by our numerical calculations,
see below.

The scattering current dipole in the considered scheme is induced by
magnetically coupled flux qubit. Its amplitude is: $\pm\mu = \pm I_p
M / 2L_{cl}J_c\lambda_J$, where $\pm I_p$ is the persistent current
circulating in the qubit (the sign corresponds to the current
direction), $M$ is the mutual inductance between the qubit and the
coupling loop, $L_{cl}$ is the inductance of the coupling loop.
According to results of the existent experimental works \cite{F} the
values of the dipole amplitude is about $\mu = 0.1$.

\subsection{Optimization of the fluxon dynamics.}

We start consideration of the measurement process based on soliton
scattering from study of detector response dependences on the
starting fluxon velocity and its stationary velocity. The current
dipole amplitude and its width are chosen to be $\mu = 0.1$ and $d =
20$, respectively. The JTL length is $L = 3 d = 60$. The dipole is
placed at the center of the JTL $x_{dc} = L/2 = 30$. At the first
step we consider the case where the starting fluxon velocity is
equal to the stationary one $u(t_0) = u_{st}$. The dipole polarity
is marked as ``positive'' - ``$+\mu$'' for the case where its first
pole is co-directed with the bias current (the first pole
accelerates the fluxon) and ``negative'' - ``$-\mu$'' otherwise (the
first pole decreases the bias current and decelerates the fluxon).
The damping is assumed to be vanishing ($\alpha~\rightarrow~0$).

\begin{figure}[t]
\resizebox{1\columnwidth}{!}{
\includegraphics{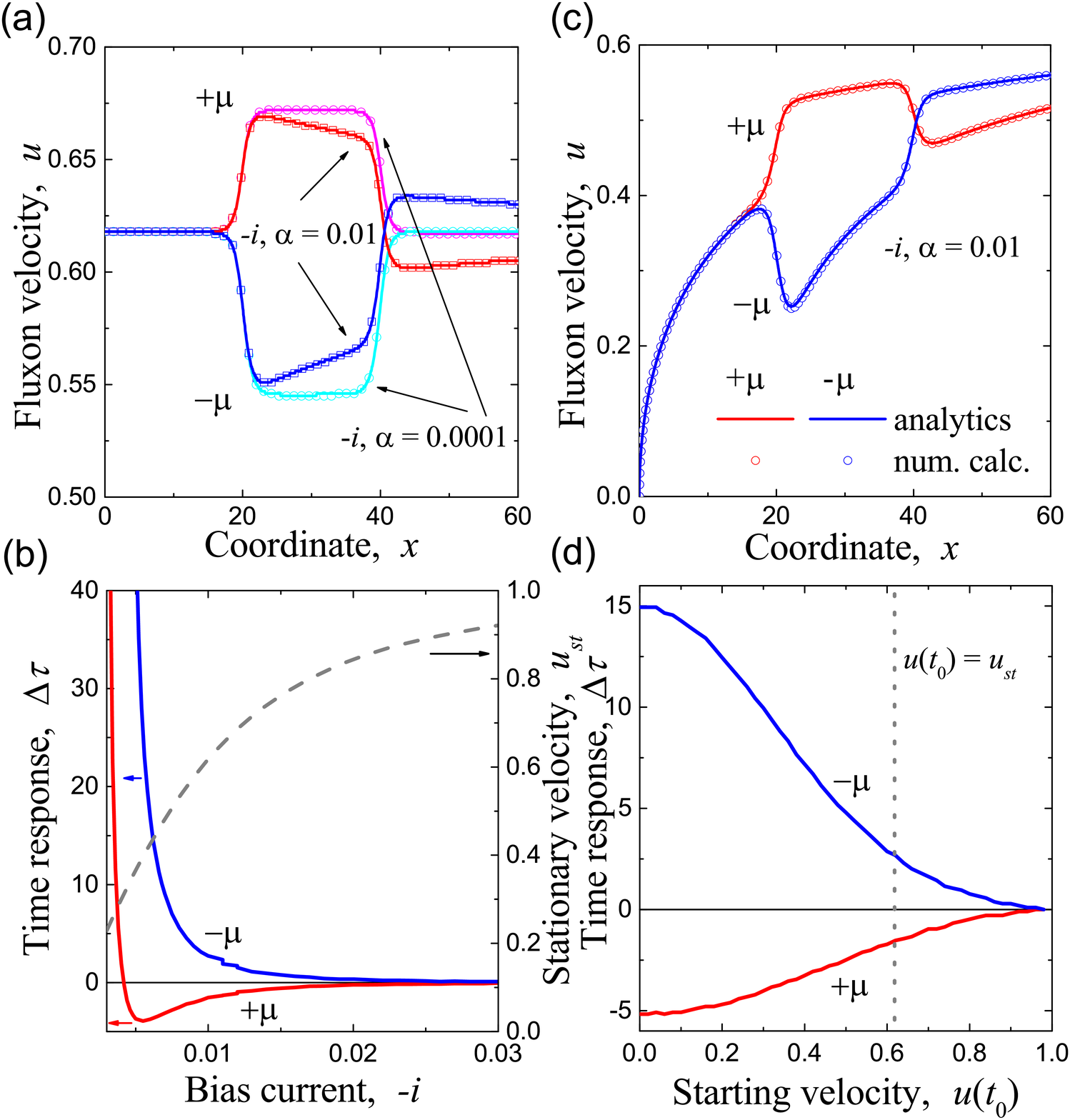}}
\caption{The fluxon velocity dependences on the coordinate for the
both dipole polarities calculated using the presented analytical
approach (solid curves) and numerically, using the system
(\ref{duX}) (dotes), with fluxon starting velocity equal to the
stationary one $u(t_0) = u_{st}$ (a), and equal to zero $u(t_0) = 0$
(c). (b) The detector time responses versus the bias current for
$u(t_0) = u_{st}$ ($u_{st}$ is defined by (\ref{ust}) and shown by
the dashed curve); $\alpha = 0.01$. (d) The time responses versus
the fluxon starting velocity for $-i,\alpha = 0.01$; the vertical
line shows the stationary velocity. The JTL and the dipole
parameters are: $L = 60$, $d = 20$, $x_{dc} = 30$, $\mu = 0.1$,
$\gamma = 0$.} \label{Fig3}
\end{figure}

Fluxon scatterings at the dipole poles provide deviation of the
fluxon velocity from the stationary one $\Delta u = u - u_{st}$
while fluxon moves inside the dipole, that is further detected as
the time response. The dependences of the fluxon velocity on the
coordinate for the both dipole polarities obtained using the
presented analytical approach (solid lines) and numerical
calculation of the system (\ref{duX}) (dots) are shown in
Fig.~\ref{Fig3}a. The JTL parameters for simulation of the ballistic
regime are as follows: $-i = 0.0001$, $\alpha = 0.0001$, $\gamma =
0$. It is seen that the data obtained with the both (analytical and
numerical) approaches are consistent perfectly.

Note, that fluxon deceleration can be more pronounced than
acceleration due to relativistic dependence of the effective fluxon
mass on its velocity. The detector response for the negative dipole
polarity can be greater than for the positive one, accordingly.

To take into account the fluxon velocity relaxation we calculate the
same velocity curves for realistic parameters $-i,\alpha = 0.01$,
see Fig.~\ref{Fig3}a. Since the fluxon velocity becomes closer to
the stationary value after the first scattering, the second
scattering provides an extra compensation of the velocity deviation,
so the deviation changes its sign. Thus, the relaxation serves for
decrease of the detector response with the damping increase.

Fig.~\ref{Fig3}b shows the detector time response $\Delta\tau =
\tau_\mu - \tau_0$ (the index $\mu$/$0$ represents the
presence/absence of the scattering) calculated numerically for the
same damping but for the different bias current values determining
the stationary (and the starting) fluxon velocity as it follows from
the expression (\ref{ust}). Small velocity corresponds to small
effective mass that makes a fluxon more susceptible to the
scattering effect increasing the response. Rapid increase of the
response with the bias current decrease indicates existence of the
threshold bias current. This threshold current corresponds to fluxon
capturing by the first (or the second) dipole pole in the case of
the negative (the positive) dipole polarity.

It is seen that the bias current maximizing the difference between the
responses for the opposite dipole polarities is in the vicinity of
the threshold current for the negative dipole. However, this current
is impractical for implementation of series measurements. For
reliable series detection (which is required for quantum
measurements) one needs to shift the working bias current upward to
guarantee that fluctuations will not cause the fluxon capturing.
Still, there is a possibility to increase the response by tuning the
starting fluxon velocity.

Fig.~\ref{Fig3}c shows the fluxon velocity curves for the same
values of the bias current and the damping ($-i,\alpha = 0.01$) as
the ones taken for calculation of corresponding curves for
nonvanishing damping shown in Fig.~\ref{Fig3}a but for zero starting
velocity $u(t_0) = 0$. Fig.~\ref{Fig3}d presents the dependences of
the detector time response on the starting velocity for these JTL
parameters calculated numerically using the system (\ref{duX}). The
decrease of the starting velocity in our case can lead to 5 times
increase in difference between the time responses for the opposite
$\mu$. We should note, that difference of signs of the time
responses for the opposite dipole polarities can provide an
advantage for a measurement scheme which uses digital comparator at
the output.

\subsection{SNR of the fluxonic ballistic detector.}

In the work \cite{FSSK} it was argued that the major sources of the
measurement errors in the fluxonic detector are the fluxon
propagation time jitter due to the thermal fluctuations and the
intrinsic qubit relaxation. If the time of the measurements is much
smaller than the qubit relaxation time, then the detector SNR can be
calculated according to the equation (\ref{SNR}). The total jitter
$\sigma_\Sigma = \sqrt{\sigma_\mu^2 + \sigma_0^2}$ in this equation
is the standard deviation of the detector time response $\Delta\tau$
which can be obtained using the jitters corresponding to the fluxon
propagations through the JTL with ($\sigma_\mu$) and without
($\sigma_0$) scattering.

\begin{figure}[t]
\resizebox{1\columnwidth}{!}{
\includegraphics{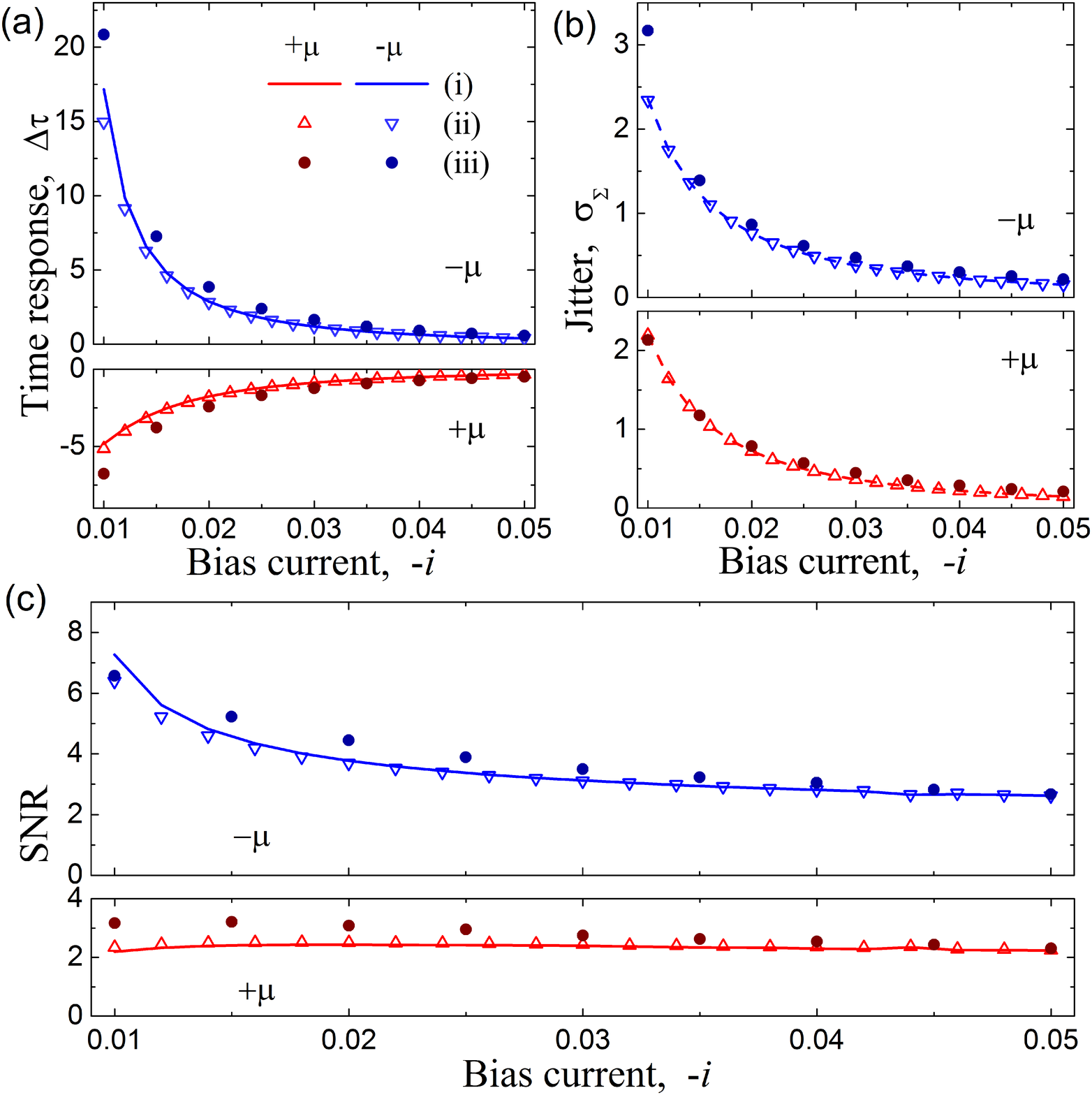}}
\caption{The detector time response (a), the total jitter (b) and
the SNR (c) versus the bias current for the both dipole polarities
evaluated using (i) the equations (\ref{uX0})-(\ref{SNR}) (solid
curves), (ii) numerical calculations of the system (\ref{duX}) with
$\xi = 0$ and further accounting for the noise effect using the
equations (\ref{D})-(\ref{SNR}) (open triangles), and (iii)
numerical calculations of the SG equation (\ref{SG}) with averaging
over ensemble of 10000 realizations (filled dots). Legend for all
the panels is shown in the panel (a). $L = 60$, $d = 20$, $x_{dc} =
30$, $-i,\alpha = 0.01$, $\gamma = 10^{-3}$, $\mu = 0.1$, $u(t_0) =
0$.} \label{Fig4}
\end{figure}

To verify our theoretical approaches for evaluation of the detector
parameters we calculate the detector time response, the total jitter
and the SNR versus the bias current for the both dipole polarities
using the three presented methods: (i) the analytical one (equations
(\ref{uX0}) - (\ref{SNR})), (ii) numerical calculation of the system
(\ref{duX}) with $\xi = 0$ and further accounting for the noise
effect using equations (\ref{D}) - (\ref{SNR}), and (iii) numerical
calculation of the SG equation (\ref{SG}) with averaging over
ensemble of 10000 realizations. In case (iii) the simulations have been performed using the
original implicit finite-difference scheme \cite{Fed,Fed2}, which is similar
to the Crank-Nicolson one, but with the account of the white noise source.
The JTL and the dipole parameters
are the same as before: $L = 60$, $d = 20$, $x_{dc} = 30$,
$-i,\alpha = 0.01$, $\mu = 0.1$, but with $\gamma = 10^{-3}$. The
starting fluxon velocity is equal to zero $u(t_0) = 0$. The results
are shown in Fig.~\ref{Fig4}.

It is seen that the data obtained using the all three methods are
well consistent. Some discrepancy occurs in the range of the small
bias current values where the scattering effect is especially
highlighted.

The jitter increase with the bias current (and the stationary
velocity) decrease can be qualitatively explained by relativistic
decrease of the effective fluxon mass that makes fluxon dynamics
more affected by noise. Despite this increase, for the negative
dipole polarity the SNR still grows a bit toward the small bias
current values (see Fig.~\ref{Fig4}c) because of more rapid increase
of the time response. Contrary to this, for the positive dipole
polarity the SNR curve is nearly flat because of limited growth of
the time response in this case, see also Fig.~\ref{Fig3}b.

\begin{figure}[t]
\resizebox{1\columnwidth}{!}{
\includegraphics{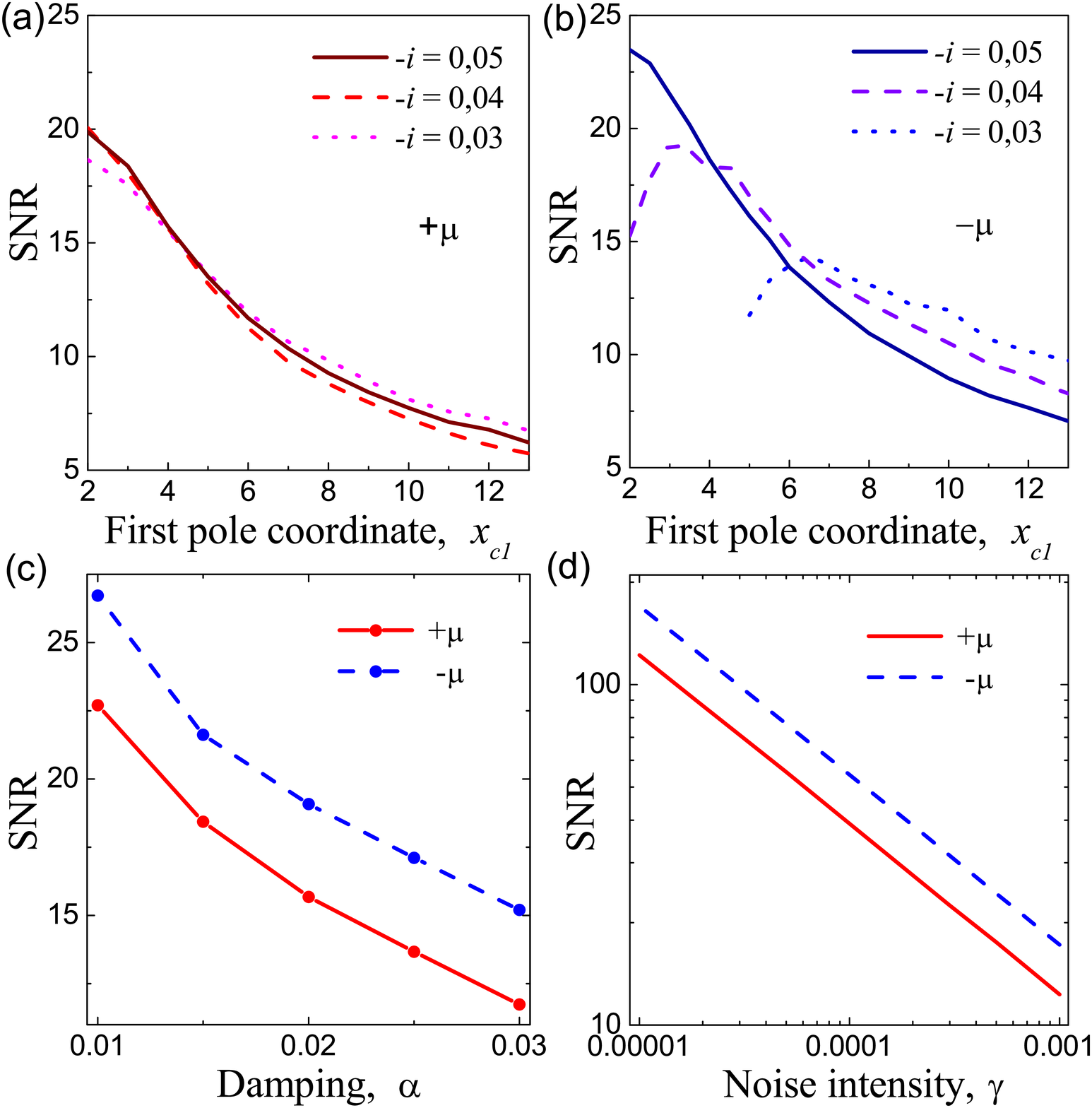}}
\caption{The detector SNR versus the first dipole pole position
$x_{c1}$ in the JTL for the positive (a) and the negative (b) dipole
polarities, and the bias current values $-i = 0.03,~0.04,~0.05$;
$x_{c2} = 55$, $\alpha = 0.02$, $\gamma = 10^{-3}$. (c) The SNR
versus the damping for the optimum bias current $-i = 0.048$ at
$x_{c1} = 5$. (d) The SNR versus the noise intensity for the same
parameters and $\alpha = 0.02$. $L = 60$, $\mu = 0.2$, $u(t_0) =
0$.} \label{Fig5}
\end{figure}

Along with the quite smooth dependences of the SNR on the bias
current values corresponding to a wide range of the stationary
fluxon velocities, we find more pronounced SNR dependence on the
instant fluxon velocity before the scattering at the first dipole
pole. Since the scattering at the second pole ends formation of the
time response, we shift the second dipole pole nearly to the end of
the JTL: $x_{c2} = L - 5 = 55$. To make our results more relevant to
the experiment we increase the damping value $\alpha = 0.02$ as well
as the dipole amplitude $\mu = 0.2$ while the noise intensity and
the starting fluxon velocity are hold the same $\gamma = 10^{-3}$,
$u(t_0) = 0$. The dependences of the SNR on the first dipole pole
position for different bias currents were calculated numerically
using the SG equation (\ref{SG}). The results for the positive and
the negative dipole polarities are shown in Fig.s~\ref{Fig5}a,b
respectively.

Shift of the first dipole pole to the beginning of the JTL leads to
decrease of the instant fluxon velocity and the corresponding
decrease of the instant effective fluxon mass before the first
scattering. Since this scattering mainly forms the time response,
the SNR for the positive dipole polarity monotonically grows toward
the smaller coordinate values. At the same time, the bends of the
SNR curves for the negative dipole illustrates an increase of the
jitter impact on the SNR in the vicinity of the threshold bias
current. The tops of these bends correspond to the optimum sets of
parameters for the considered measurement procedure.

Assuming that the closest location of the first dipole pole to the
beginning of the JTL can be about $x_{c1} = 5$, we calculated the
SNR versus the damping for the optimum bias current value $-i =
0.048$ corresponding to this $x_{c1}$, see Fig.~\ref{Fig5}c. It is
seen that fluxon velocity relaxation provides nearly the same effect
on the SNR for the both dipole polarities. Finally, to evaluate the
expected SNR of the detector in the experiment, we calculated the
SNR versus the noise intensity for the damping value $\alpha = 0.02$
and the same bias current. The results are presented in
Fig.~\ref{Fig5}d. According to our assumptions, the SNR scales as
$SNR \sim \sigma_\Sigma^{-1} \sim \gamma^{-1/2}$ that is consistent
with the results of the work \cite{SKBPK}. For the estimated noise
intensity $\gamma = 10^{-5}$ the SNR is above 100.

\section{Conclusion}
In conclusion, we have developed analytical approach for calculation
of relativistic dynamics of soliton scattering at weak short
inhomogeneity of the driving force and account for the presence of
the thermal fluctuations. We have generalized this approach for an
arbitrary number of inhomogeneities as well as considered numerical
approaches for calculation of the dynamics for arbitrary parameters
of the system. We have considered the scattering as a measurement
procedure by example of the fluxonic ballistic detector, exploiting
the developed methods for its optimization. The negative role of the
damping in the system for formation of the detector time response
was outlined as well as using of accelerated fluxon motion leading
to increase of the response was argued. Finally, we have optimized
the measurement scheme configuration for experimentally relevant
parameters and obtained the SNR value above 100. Since the obtained
time response and its standard deviation in the frame of the
detector model (taking into account only fluxon dynamics in the
JTLs) are by an order of magnitudes larger than the time resolution
and the thermal jitter of digital superconducting delay detector
\cite{FSSK, GP}, this SNR reflects quantitatively correct estimation
of attainable performance of the measurement process which is
planned to be realized experimentally.

\section{Acknowledgements}

This work was supported by Ministry of Education and Science of the
Russian Federation, grant Nos. 14.Y26.31.0007, 3.2054.2014/K,
RFMEFI58714X0006, RFBR projects 14-02-31002-mol$\_$a,
15-32-20362-mol$\_$a$\_$ved, 15-02-05869, Russian President grant
MK-1841.2014.2, Dynasty Foundation, and in the framework of Increase
Competitiveness Program of Lobachevsky NNSU under contract no.
02.B.49.21.0003.

\appendix
\section{Calculation of soliton scattering dynamics}
The perfect SG equation
\begin{equation}
{\phi}_{tt}-{\phi}_{xx}+\sin(\phi)=0 \label{SGp}
\end{equation}
can be written as a Hamiltonian system for $(\phi,\phi_t)$ with the
Hamiltonian
\begin{equation}
H^{SG} = \int_{-\infty}^\infty \left(\tfrac{1}{2}\phi^2_t +
\tfrac{1}{2}\phi^2_x + 1 - \cos(\phi)\right)dx. \label{HSGp}
\end{equation}
This system supports soliton solution which can be analytically
presented by the two-parameter formula
\begin{equation}
\phi_0(x,t;x_0,u) = 4\tan^{-1}\left[\exp\pm\left(\frac{x - ut -
x_0}{\sqrt{1-u^2}}\right) \right], \label{Phi0}
\end{equation}
where the velocity $|u| < 1$ and the coordinate $x_0$ are the
parameters, and $\pm$ represents soliton or antisoliton state.

According to the collective coordinate perturbation analysis performed
by McLaughlin and Scott \cite{McS} we consider soliton dynamics in a
real physical system by introducing weak structural perturbation
into the perfect SG equation in the form
\begin{equation}
{\phi}_{tt}-{\phi}_{xx}+\sin(\phi) = \epsilon f, \label{SGpp}
\end{equation}
where $0\leq|\epsilon|\ll1$, and find the response of the SG wave
solution $\vec{W} = {\phi \choose \phi_t}$ (it is assumed that
initially the wave is precisely the pure soliton state $\vec{W}_0 =
{\phi_0 \choose \phi_{0t}}$ corresponding to (\ref{SGp})) to this
perturbation in the form
\begin{equation}
\vec{W} = \vec{W}_0 + \epsilon\vec{\mathfrak{w}}, \label{SGpp}
\end{equation}
by establishing equations for $\vec{\mathfrak{w}}$ governing
modulation of the wave parameters in time.

For the considered case of a single soliton wave (\ref{Phi0}) these
equations are as follows\cite{McS}:
\begin{subequations} \label{dudx0}
\begin{equation}
\frac{du}{dt} = \mp\epsilon\tfrac{1}{4}(1-u^2)\int_{-\infty}^\infty
f[\phi_0(\Theta)]\sech(\Theta)dx,
\end{equation}
\begin{equation}
\frac{dx_0}{dt} =-\epsilon\tfrac{1}{4}u(1-u^2)\int_{-\infty}^\infty
f[\phi_0(\Theta)]\Theta\sech(\Theta)dx,
\end{equation}
\end{subequations}
where $\Theta(x,t) = \left(x - \int_{t_0}^t u(t')dt' -
x_0\right)/\sqrt{1-u^2}$. Note, that the system (\ref{dudx0}) can be
obtained just from the energy equation and the equation for the
soliton momentum $p = -\frac{1}{8}\int_{-\infty}^\infty \phi_{0x}
\phi_{0t} dx$ correspondingly:
\begin{subequations} \label{dEdp}
\begin{equation}
\frac{dH^{SG}(\phi_0)}{du}\frac{du}{dt} = \epsilon
\int_{-\infty}^\infty f(\phi_0)\phi_{0t}dx,
\end{equation}
\begin{equation}
\left(\frac{d}{du}\int_{-\infty}^\infty \phi_{0x}\phi_{0t}
dx\right)\frac{dx_0}{dt} = \epsilon \int_{-\infty}^\infty
f(\phi_0)\phi_{0u}dx.
\end{equation}
\end{subequations}

Defining the central soliton coordinate as
\begin{equation}
X \equiv \int_{t_0}^t u(t')dt' + x_0,
\end{equation}
so that $\dot{X} = u + \dot{x}_0$, and substituting the right-hand
side of the considered SG equation (\ref{SG}) for $\epsilon f$ in
the system (\ref{dudx0}) one directly obtains the system
(\ref{duX}).

Since for $\mu, \xi = 0$ the perturbation $\epsilon f$ is the even
function of $\Theta$, $\dot{x}_0 = 0$ and $\dot{X} = u$ as it
follows from (\ref{dudx0}b). In this case the equation
(\ref{dudx0}a) has a simple form
\begin{equation}
\dot{p} =-\alpha p - \pi i/4.
\end{equation}
Its solution is shown by expression (\ref{p}) (corresponding $u(t)$,
$X(t)$ dependences are the equations (\ref{uX0})).

For the ballistic regime $i, \alpha, \xi = 0$, the right-hand side
of the system (\ref{duX}) contains only terms with $\theta$ and
therefore it is more convenient to seek for $u$, $X$ dependences on
this argument modifying the equations (\ref{duX}) correspondingly:
\begin{subequations} \label{dudXscs}
\begin{equation}
\frac{du}{d\theta} =
-\frac{\sqrt{1-u^2}}{u}\frac{\frac{\mu}{4}(1-u^2)\sech(\theta)}{1 -
\frac{\mu}{2}\sqrt{1-u^2}\theta\sech(\theta)},
\end{equation}
\begin{equation}
\frac{dX}{d\theta} = \frac{\sqrt{1-u^2}}{u}\frac{u -
\frac{\mu}{4}u(1-u^2)\theta\sech(\theta)}{1 -
\frac{\mu}{2}\sqrt{1-u^2}\theta\sech(\theta)}.
\end{equation}
\end{subequations}
At the moment of the scattering $\theta = 0$ one can make
approximation for the denominator of the equations (\ref{dudXscs})
\begin{equation}
\left(1 - \frac{\mu}{2}\sqrt{1-u^2}\theta\sech(\theta)\right)^{-1}
\approx 1 + \frac{\mu}{2}\sqrt{1-u^2}\theta\sech(\theta)
\end{equation}
and holding only the terms containing $\mu$ in the first power we
obtain the solution (\ref{uX1}). Here we consider only forward
scattering in accordance with our assumption $|\mu| \ll 1$.

The velocity shift (\ref{ustep}) provided by the scattering for
incident soliton is obtained from conditions at infinity (for $u_0
> 0$: $\theta = -\infty,~c_{-\infty} = \pi/4\rightarrow \theta = +\infty,~c_{+\infty} =
-\pi/4$, and vice versa for $u_0 < 0$). This shift (\ref{ustep}) is
used for definition of the new starting conditions for soliton
velocity relaxation process that should be applied at the moment of
the scattering $t^{sc}$.

In the case of arbitrary number of scatterings $N$, one can
successively find the moments of the scatterings $t_n^{sc}$ ($n =
1\ldots N$) (and corresponding starting conditions) considering the
velocity relaxation process of the soliton scattered at the $n-1$
inhomogeneity as incident soliton dynamics for $n$-th scattering.
The $u^{sc}_n(t)$ dependence can then be constructed iteratively as
follows:
\begin{equation}
u^{sc}_n(t) = \begin{cases} u^{sc}\left(t < t^{sc};~u_0 =
u^{sc}_{n-1}(t), \theta =\frac{X^{sc}_{n-1}(t) - x_{cn}}{\sqrt{1 -
(u^{sc}_{n-1}(t))^2}}\right),\\
u^{sc}\left(t \geq t^{sc};~u_0 = u^{rel}(t), \theta =
\frac{X^{rel}(t) - x_{cn}}{\sqrt{1 - (u^{rel}(t))^2}}\right),
\end{cases}
\end{equation}
where $u^{sc}_0(t) = u^{rel}(t;~t_0,u^{rel}(t_0))$. The
$X^{sc}_n(t)$ dependence can be obtained similarly.

\end{document}